\documentclass[conference]{IEEEtran}
\IEEEoverridecommandlockouts
\usepackage{cite}
\usepackage{algorithmic}
\usepackage{graphicx}
\usepackage{textcomp}
\usepackage{xcolor}
\def\BibTeX{{\rm B\kern-.05em{\sc i\kern-.025em b}\kern-.08em
    T\kern-.1667em\lower.7ex\hbox{E}\kern-.125emX}}
\begin{document}

\title{Convectively-coupled High-frequency Atmospheric waves triggered Kerala floods in 2018 and 2019\\

}

\author{\IEEEauthorblockN{Kiran S. R.}
\IEEEauthorblockA{\textit{Department of Civil Engineering} \\
\textit{Central Polytechnic College}\\
Thiruvananthapuram, Kerala\\
Email: kiransreekumarr@gmail.com\\
ORCID Id: 0000-0003-3200-8624}
}

\maketitle

\begin{abstract}
Floods have repeatedly battered the South Indian state, Kerala, as a result of the unprecedented heavy rainfall during Boreal Summers, in recent years. The state witnessed large departures from normal rainfall in 2018 and 2019. Previous studies have seldom adopted a systematic approach to understand the phenomenon responsible for the recurrent extreme events. Hence, this study, based on spectral methods, identifies a characteristic propagation of high-frequency equatorial waves in the atmosphere, which travelled from near tropical west Pacific to the east coast of Africa. These waves stimulated intense convection and ensured sufficient availability of moisture over the state, and are hence responsible for Kerala Floods. 
\end{abstract}


\section{Introduction}
Kerala, the south-west coastal state of India, experienced heavy rainfall and consequent floods in 2018 and 2019 during Summer. The state received extremely heavy rainfall during the South-West Monsoon in the two years --  from $8^{th}$ to $17^{th}$ August 2018  and from $3^{rd}$ to $10^{th}$ August 2019, which caused unprecedented floods resulting in massive loss of life and property. These recurrent extreme events in the state have raised questions of changing climate in the southern part of Peninsular India. Hence, there arose a necessity to identify the physical phenomenon which derailed the hydrology of the State in recent years. 

Vimal and Harsh (2019) observed no increase in mean and extreme precipitation in Kerala over the past six decades and identified that the extreme rainfall event in August 2018 was driven by anomalous atmospheric conditions due to climate variability rather than anthropogenic factors \cite{b1}. Hunt and Menon (2020) concluded that the precipitation in Kerala may be 18\% heavier for the pre-industrial period and 36\% heavier for a prospective climate in 2100 \cite{b2}. According to Ref. \cite{b3}, the presence of a strong westerly jet along with the formation of the offshore vortex trough and the transport of mid-tropospheric moisture from various sources under the presence of conducive vertical shear of horizontal wind may have contributed to the extreme rainfall over Kerala in 2018. Further, there are theories that the strong westerly flow, which is a part of the cross equatorial Monsoon flow, was constantly propelled by a couple of cyclonic systems over Bay of Bengal, South China Sea and West Pacific \cite{b4}. Similarly, Ref.\cite{b5} deduced that the Kerala floods of 2018 was triggered by the formation of an intense and long-duration atmospheric river with more than 60\% of moisture supply from the Central-Eastern Indian Ocean. 

Although, the above studies attempted to investigate the extreme precipitation in 2018 from the general wind field conditions existed then, no conclusive evidences to support their theories exist hitherto, even after the extreme events repeated in 2019 and 2020 during the same season. This incited the necessity to investigate the climate dynamics responsible for the recurrent disaster.

\section{Data \& Methodology}
Tropical Rainfall Measuring Mission (TRMM) gives daily rainfall data at a spatial resolution of $0.25^{o}$ $\times$ $0.25^{o}$ over the Indian Ocean region. In addition, the daily gridded rainfall data from raingauge observations over Kerala is obtained from Indian Meteorological Department (IMD) from 2011 to 2019. Further, the ERA5 Reanalysis of the European Centre for Medium-Range Weather Forecasts (ECMWF) gives the specific humidity (kg/kg) and winds (zonal \& meridional winds in m/s and vertical velocity (or omega) in Pa/s) for different vertical pressure levels from 1000 to 100hPa.

\begin{figure*}
	\centering
	\hspace*{-0.5cm}
	\includegraphics[width=1.05\linewidth]{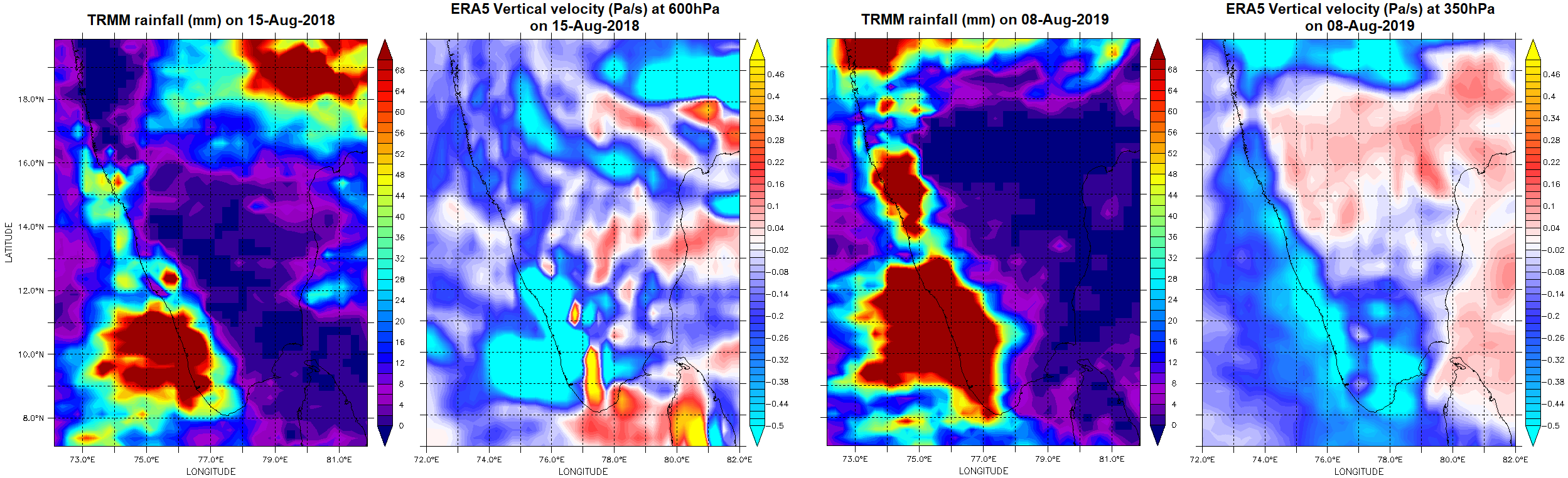}
	\caption{(first \& third figure) TRMM rainfall on the days of peak rainfall in 2018 and 2019 respectively. (second \& fourth figure) Corresponding values of ERA5 Vertical velocity at 600hPa and 350hPa respectively.}
	\label{fig:fig1}
\end{figure*}

\begin{figure}
	\centering
	\includegraphics[width=1\linewidth]{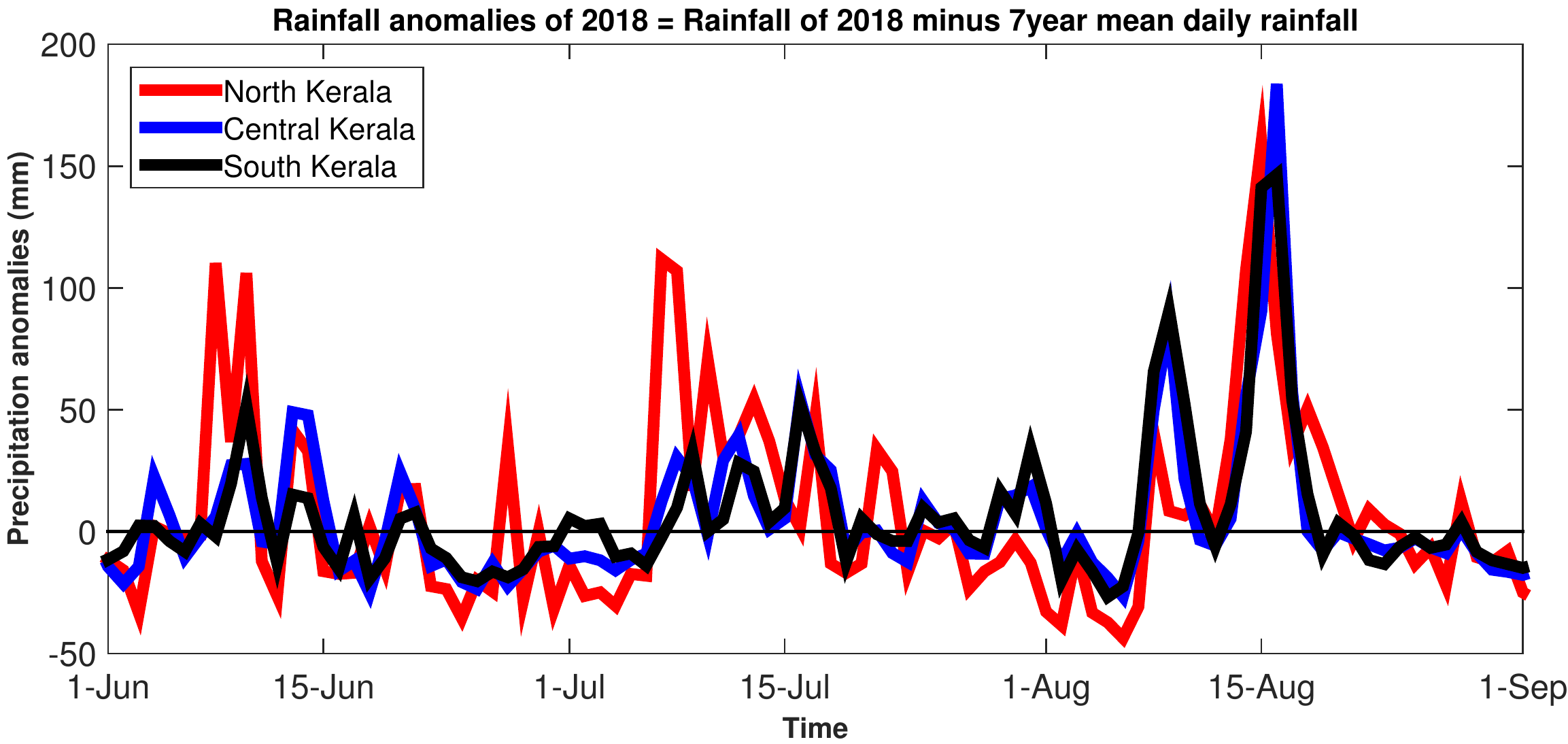}
	\caption{IMD daily rainfall anomalies (in millimetres) of 2018 over North, Central and South Kerala, measured with respect to the 7 year mean precipitation.}
	\label{fig:fig2}
\end{figure}

\begin{figure}
	\centering
	\includegraphics[width=1\linewidth]{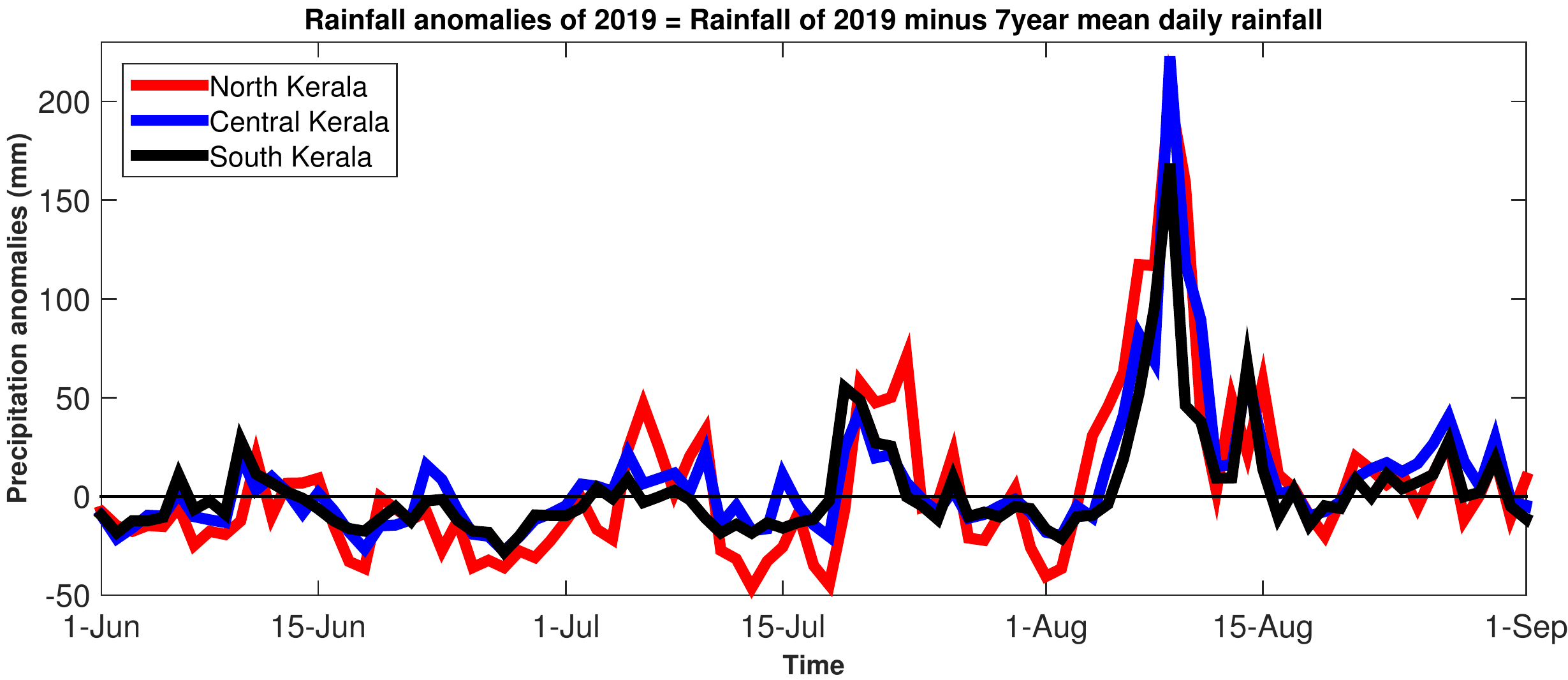}
	\caption{IMD daily rainfall anomalies (in millimetres) of 2019 over North, Central and South Kerala, measured with respect to 7 year mean precipitation.}
	\label{fig:fig3}
\end{figure}

In this study, the daily departures from the mean of atmospheric variables during 2018 and 2019 by deducting the average of daily data of the atmospheric variables from 2011 to 2017 from the daily data for the years 2018 and 2019 separately. Since, the two-day maximum precipitation of 2018 was abnormally high (220mm) and surpassed even the $95^{th}$ percentile of 70 years (150mm), a 7-year mean would suffice to capture the anomaly \cite{b1}. Hovmöller diagrams, which depict the temporal variation of atmospheric variables along zonal/meridional direction, are used to identify the propagation of waves and its characteristics. Fast Fourier Transform (FFT) is performed on time series of atmospheric variables or its anomalies for the ease of representation in frequency domain. The dominant harmonics, thus identified, are filtered out using Band-Pass Filters (BPF) for further analysis.

\section{Results \& Discussions}
\subsection{Rainfall over Kerala in 2018 $\&$ 2019}
Kerala experienced anomalous rainfall events in 2018 and 2019, with peak daily precipitation on $15^{th}$ August and $8^{th}$ August respectively \cite{b3,b6}. Figure~\ref{fig:fig1} (first \& third figure) shows the spatial distribution of TRMM rainfall over Peninsular India on these days. One observes heavy precipitation confined only to the windward side of Western Ghats, especially its southern region. Any propensities of localized convection over the state may be determined from ERA5 omega distribution. At a geographical location in Kerala ($76.5^{o}$E, $10^{o}$N), unusually high negative values of omega (rising air) were observed at 600hPa in 2018 and at 350hPa in 2019 (Fig.~\ref{fig:fig1} (second \& fourth figure), and are hence juxtaposed with rainfall for comparison. It is interesting to conclude very high spatial correlation of precipitation with mid-tropospheric convection in 2018 and with upper-tropospheric convection in 2019, on the days of extreme events.

\begin{figure}
	\centering
	\includegraphics[width=1\linewidth]{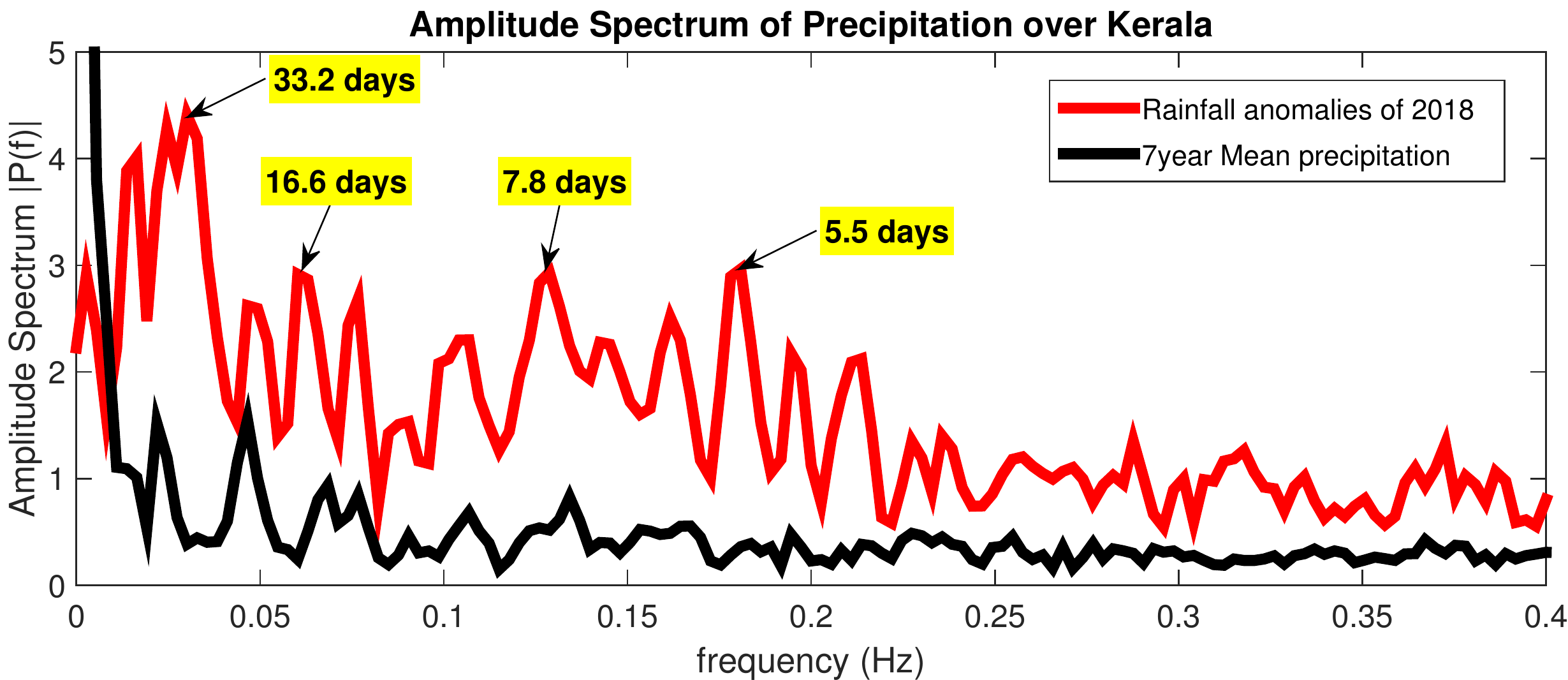}
	\caption{Amplitude spectrum of IMD rainfall anomalies of 2018 averaged over the state of Kerala, in comparison to that of the 7 year mean precipitation.}
	\label{fig:fig4}
\end{figure}

\begin{figure}
	\centering
	\includegraphics[width=1\linewidth]{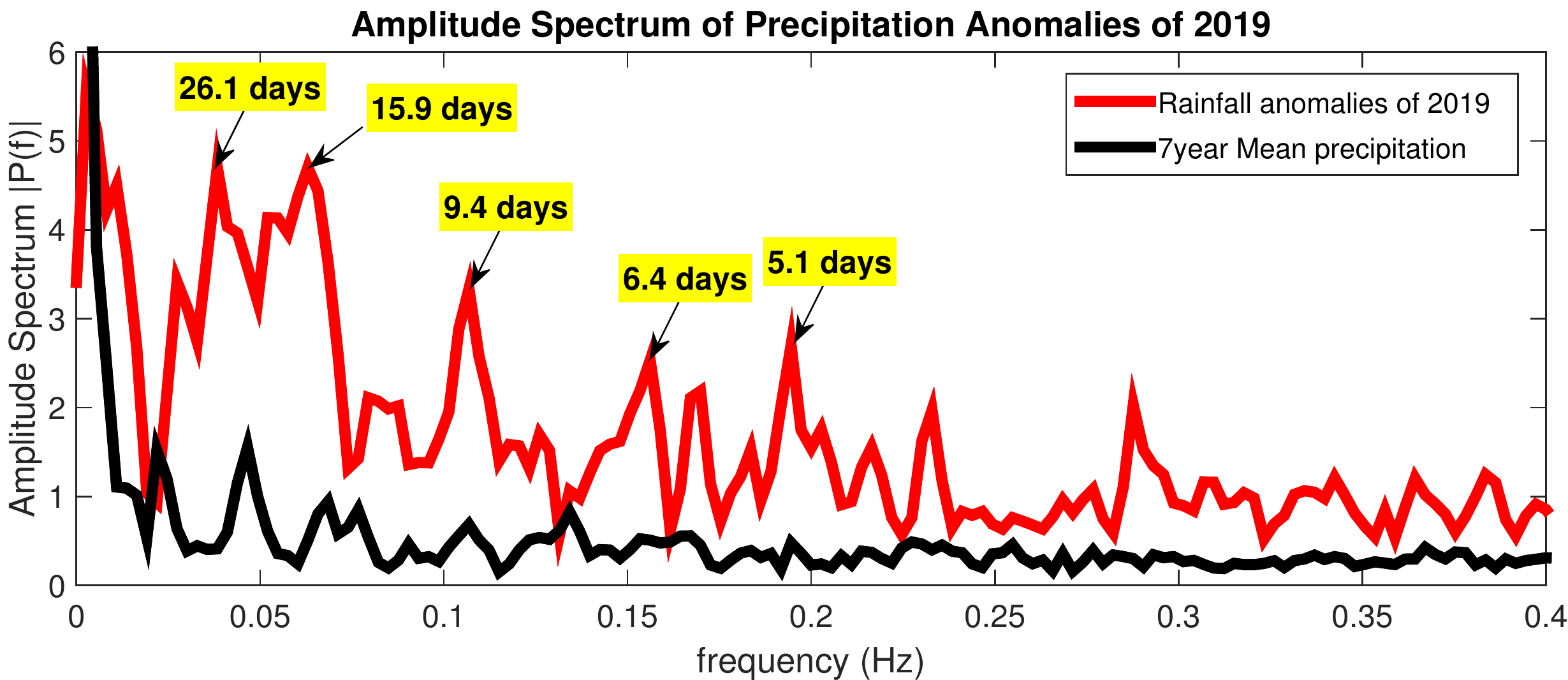}
	\caption{Amplitude spectrum of IMD rainfall anomalies of 2019 averaged over the state of Kerala, in comparison to that of the 7 year mean precipitation.}
	\label{fig:fig5}
\end{figure}

From IMD gridded rainfall point data, precipitation anomalies were determined for the North, Central and South Kerala, for the years 2018 and 2019 (Fig.~\ref{fig:fig2} and Fig.~\ref{fig:fig3}). In 2018, Kerala witnessed two distinct extreme events -- on $8^{th}$ August (over 50mm) and on $15^{th}$ August (over 150mm). Similarly, two extreme events were observed in 2019 too -- on $8^{th}$ August (over 150mm) and on $15^{th}$ August (over 50mm). 

\begin{figure}
	\centering
	\includegraphics[width=1\linewidth]{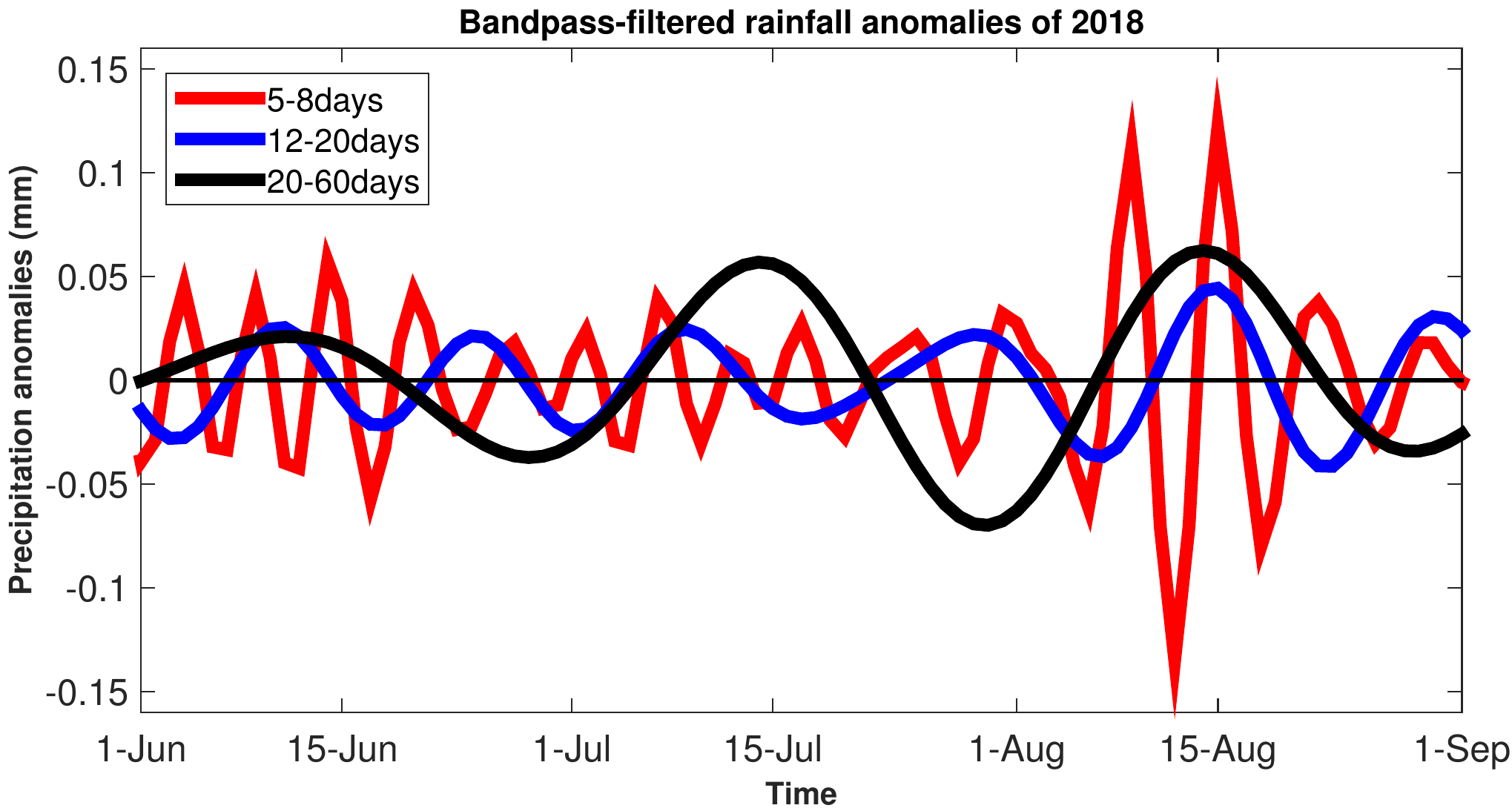}
	\caption{5-8days, 12-20days and 20-60days bandpass-filtered IMD rainfall anomalies of 2018 averaged over the state of Kerala.}
	\label{fig:fig6}
\end{figure}

\begin{figure}
	\centering
	\includegraphics[width=1\linewidth]{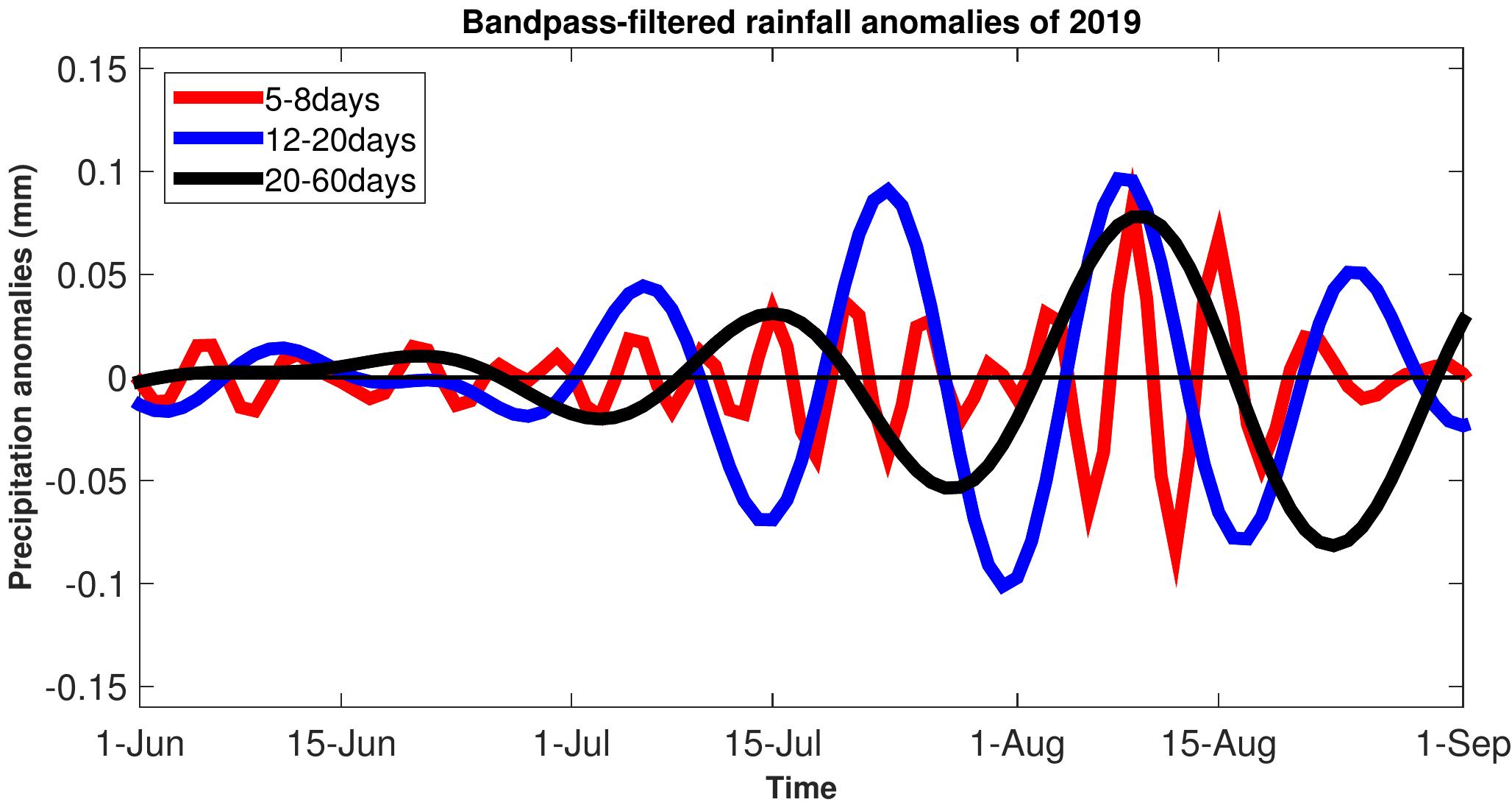}
	\caption{5-8days, 12-20days and 20-60days bandpass-filtered IMD rainfall anomalies of 2019 averaged over the state of Kerala.}
	\label{fig:fig7}
\end{figure}

\subsection{Dominant modes of variability}
Identification of the constituent harmonics, which contributed to the precipitation anomalies of 2018 and 2019, shall expose the real phenomenon responsible for the extreme event. Hence, Fast Fourier Transforms (FFT) of the spatially-averaged rainfall anomalies were performed for both the years, and the resultant amplitude spectrum are shown in Fig.~\ref{fig:fig4} and Fig.~\ref{fig:fig5}. The spectrum corresponding to the 7 year mean daily rainfall is superimposed for comparison. In 2018 and 2019, the rainfall spectra are dominated by frequencies in the bands of period 5-8 days, 12-20 days and 20-60 days. The semi-annual variability (180 days), a characteristic response of the equatorial Indian Ocean region during Spring and Fall \cite{b7}, can be neglected for this study. 

\begin{figure}
	\centering
	\includegraphics[width=1\linewidth]{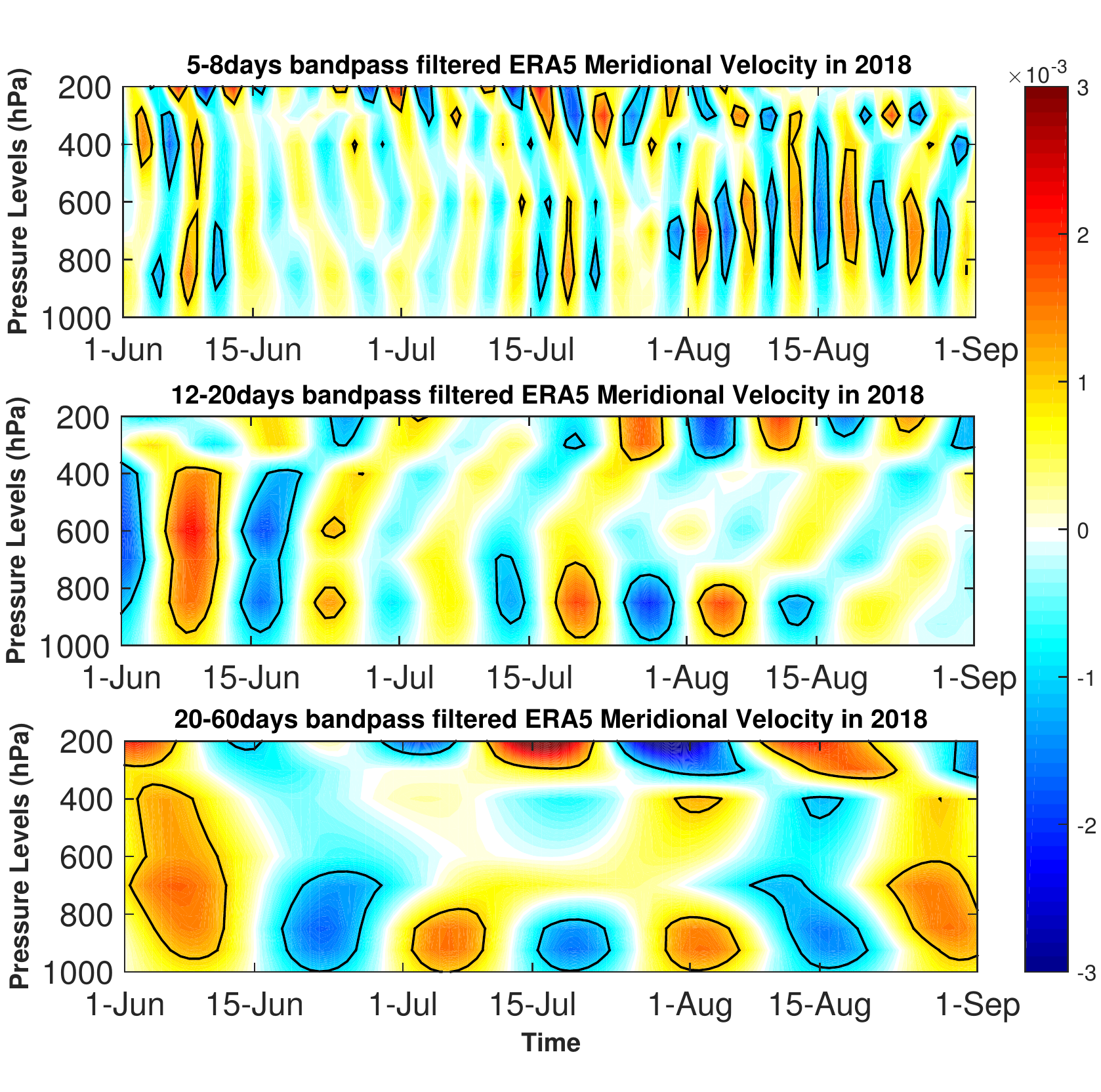}
	\caption{5-8days, 12-20days and 20-60days bandpass-filtered ERA5 Meridional velocities of 2018 averaged over the state of Kerala with altitude.}
	\label{fig:fig8}
\end{figure}

\begin{figure}
	\centering
	\vspace*{-0.5cm}
	\includegraphics[width=1\linewidth]{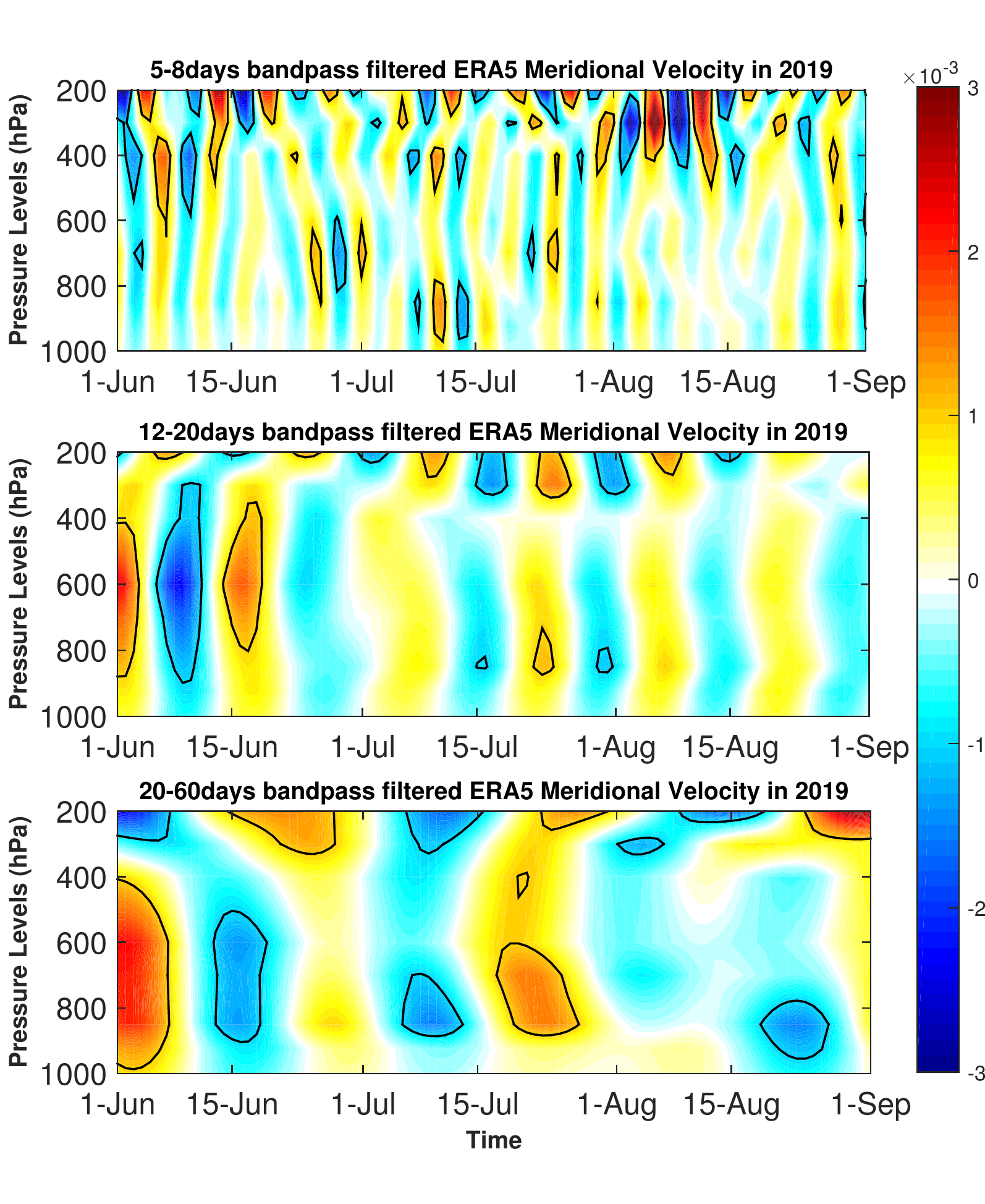}
	\caption{5-8days, 12-20days and 20-60days bandpass-filtered ERA5 Meridional velocities of 2019 averaged over the state of Kerala with altitude.}
	\label{fig:fig9}
\end{figure}


\begin{figure*}
	\centering
	\includegraphics[width=1\linewidth]{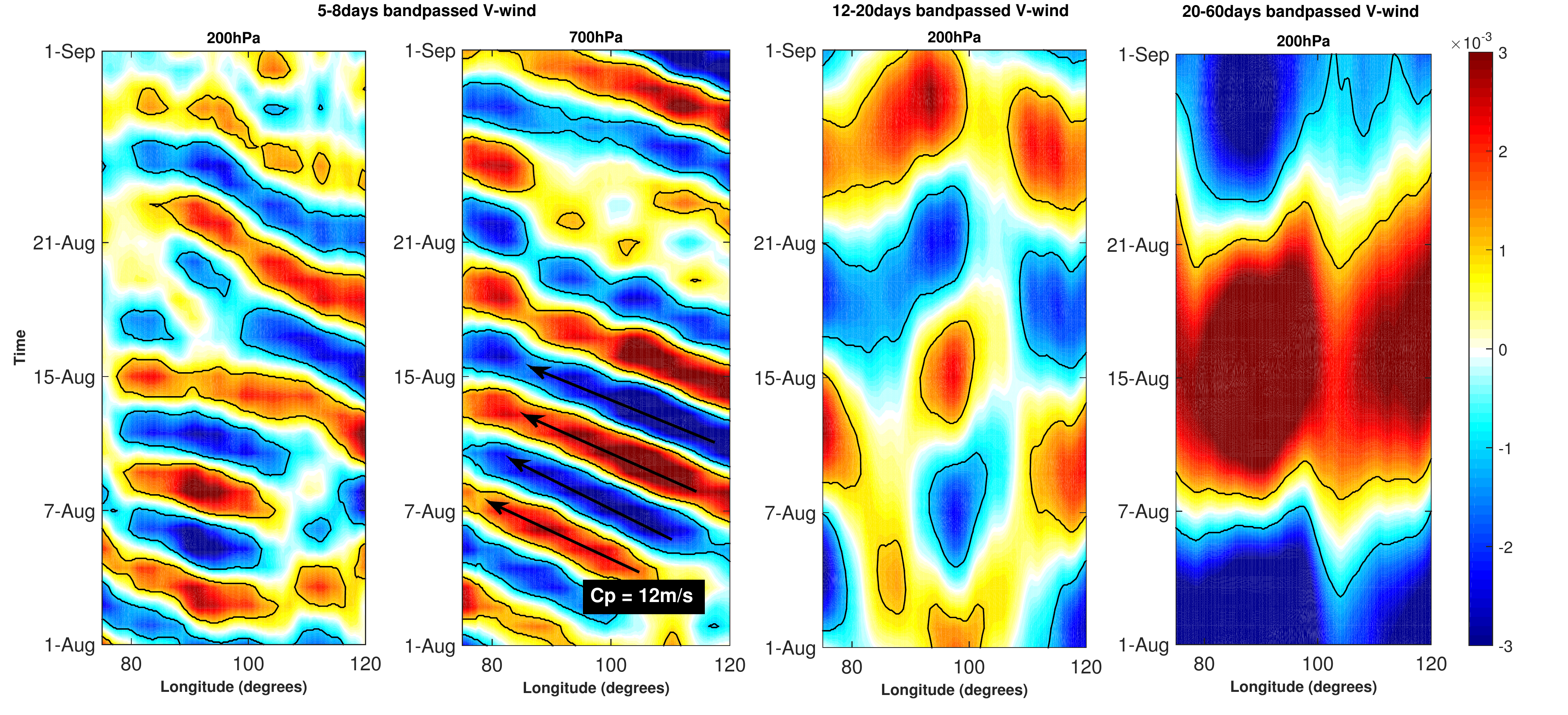}
	\caption{(first $\&$ second figure) 5-8 days bandpass-filtered ERA5 Meridional velocities of August 2018 meridionally averaged over the state of Kerala at pressure levels 200hPa and 700hPa respectively. (third figure) 12-20 days bandpass-filtered Meridional velocities at 200hPa in 2018. (fourth figure) 20-60 days bandpass-filtered Meridional velocities at 200hPa in 2018.}
	\label{fig:fig10}
\end{figure*}

For further analysis, the time series of precipitation anomalies for the concerned years were filtered separately for the bands 5-8 days, 12-20 days and 20-60 days. Fig.~\ref{fig:fig6} shows the bandpass-filtered rainfall anomalies for the year 2018. The 5-8 days high frequency mode made an exceptional contribution to the extreme event in August, compared to the other two modes. The high frequency mode rightly captured the two prominent peaks in rainfall departures during the month. In 2019, the extreme precipitation which peaked on $8^{th}$ August was the unambiguous consequence of superposition of all the three modes in-phase (Fig.~\ref{fig:fig7}). Yet, the 5-8 days periodicity leads to capture the second extreme event of the year. Hence, it is important to investigate the phenomenon which is responsible for these dominant modes of variability, especially the high frequency mode, which has assumed great significance to the extreme rainfall events of 2018 and 2019.

\begin{figure*}
	\centering
	\includegraphics[width=0.8\linewidth]{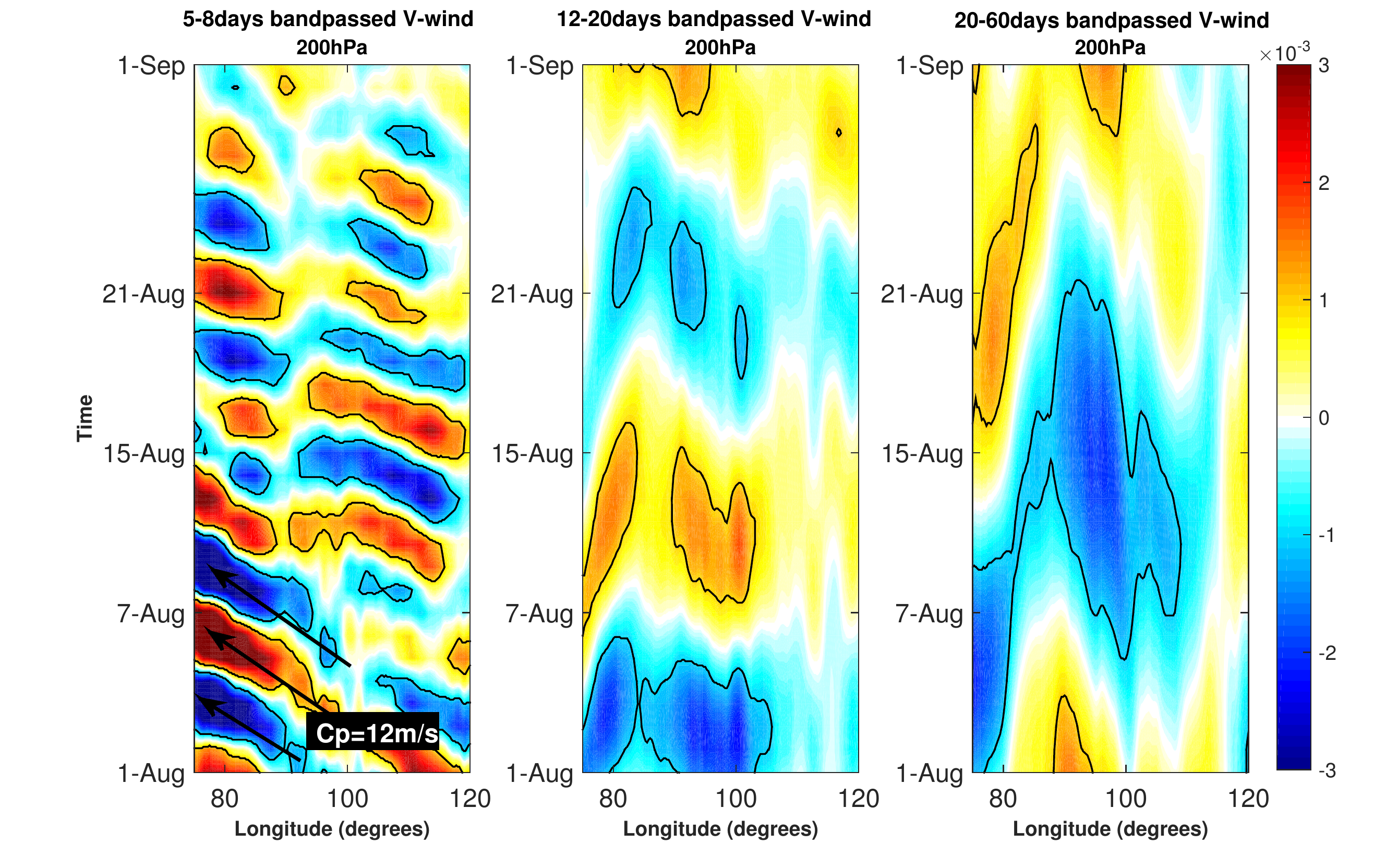}
	\caption{(first figure) 5-8 days bandpass-filtered ERA5 Meridional velocities of August 2019 averaged meridionally over the state of Kerala at 200hPa pressure level. (second figure) 12-20 days bandpass-filtered Meridional velocities at 200hPa in 2019. (third figure) 20-60 days bandpass-filtered Meridional velocities at 200hPa in 2019.}
	\label{fig:fig11}
\end{figure*}

\begin{figure}
	\centering
	\vspace*{-0.5cm}
	\includegraphics[width=0.9\linewidth]{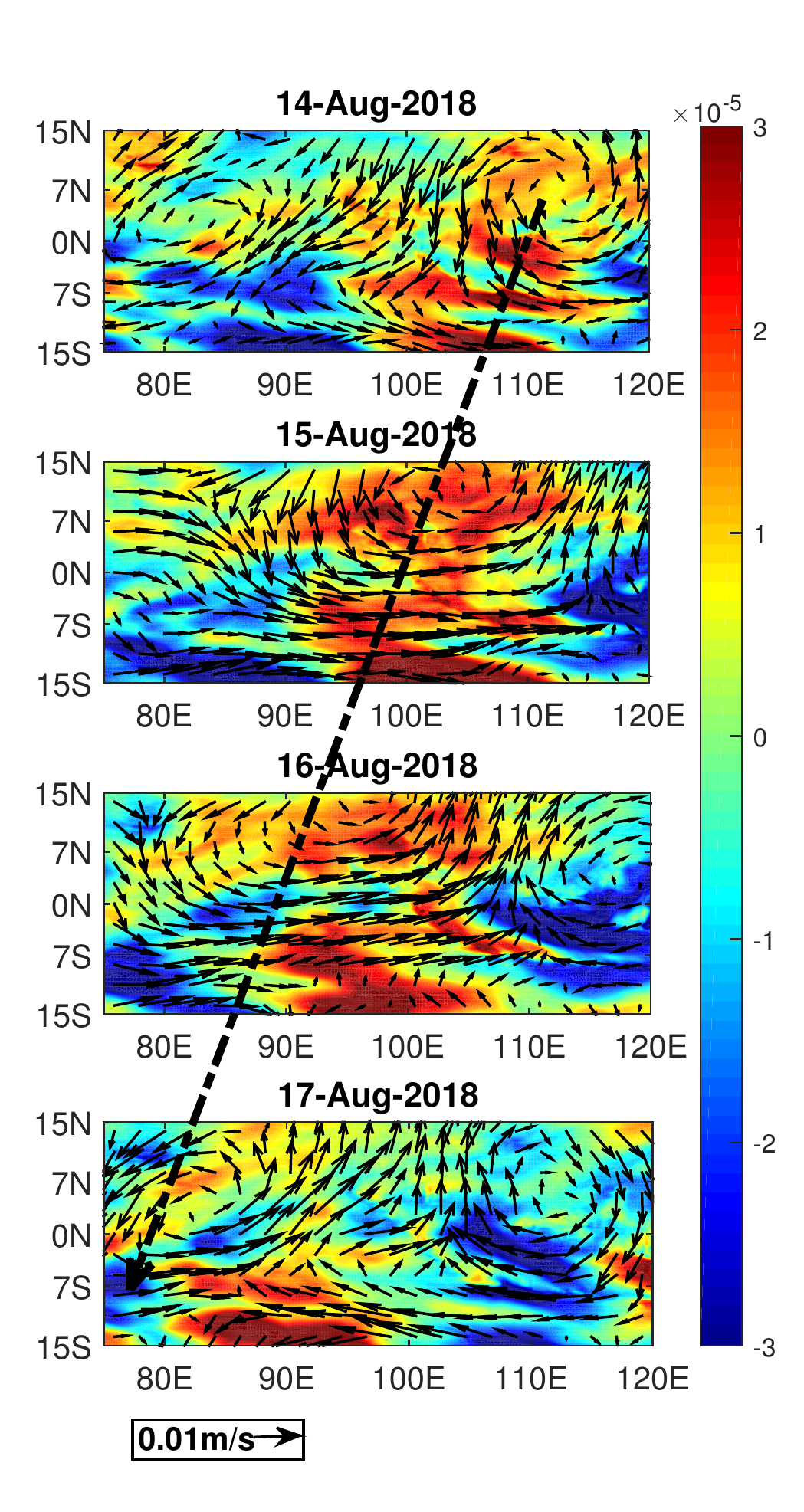}
	\caption{5-8days bandpass-filtered Specific humidity vertically integrated from surface (1000 hPa) to 600 hPa overlaid with 5-8days bandpass-filtered ERA5 winds at 700 hPa from $14^{th}$ to $17^{th}$ August 2018.}
	\label{fig:fig12}
\end{figure}

ERA5 Reanalysis meridional winds over the state of Kerala, for different pressure levels in 2018 and 2019, are bandpass-filtered for the dominant modes of variabilty (Fig.~\ref{fig:fig8} and Fig.~\ref{fig:fig9}). In 2018, the 5-8 days mode is associated with the intensification of meridional winds at mid-troposphere (800hPa - 500hPa). Similar anomalous wind patterns are seldom observed for the other modes at mid-troposphere; although these had modulated the upper troposphere, but not exclusive to period of extreme rainfall. Hence, the role of the high frequency mode in causing the Kerala floods of 2018 stands substantiated, as it conforms to the anomalous precipitation and convection observed in Fig.~\ref{fig:fig1}.

On the other hand, during the extreme event in August 2019, a stark variability in meridional winds at upper atmosphere (400hPa - 200hPa) is observed for the 5-8 days periodicity (Fig.~\ref{fig:fig9}). Other modes possessed relatively lower magnitudes at any levels during this period, although all the three modes contributed comparably to the precipitation anomalies as per Fig.~\ref{fig:fig7}. But, these modes can be observed to constitute the zonal winds over Kerala (Figure not shown), which are long deemed classical Intra-seasonal Oscillation (ISO) modes associated with the zonal component of the Monsoon Jet \cite{b11,b13}. As a consequence, it can be inferred that the Quasi-biweekly Oscillations (QBO) of 12-20 days period and the Madden-Julian Oscillations (MJO) of 20-60 days period might have exacerbated the extreme event in August 2019 \cite{b12}, but cannot drive the anomalous event by itself. Therefore, the strong connection between the high-frequency mode with the convective tendencies at mid-troposphere in 2018 and at upper-troposphere in 2019 stands established.

\subsection{Westward-propagating High-frequency Waves}
In order to study the temporal evolution of the filtered meridional winds at any pressure levels, Hovemöller diagrams are prepared for the month of August by meridional averaging over the latitudinal extend of the state of Kerala. As per the above discussion, the middle and higher levels of troposphere resonated to the 5-8 days mode during the extreme event of 2018, while only the upper atmosphere resonated to 12-20 days and 20-60 days periodicity during the same event. Therefore, Hovemöller diagram corresponding to 700hPa is prepared only for the most dominant high frequency mode, while it is prepared at 200 hPa for all the three modes of variability (Fig.~\ref{fig:fig10}). The 5-8 days mode at 700hPa exposes a interesting dynamical phenomenon, vide Fig.~\ref{fig:fig10}(second figure). During the extreme event of 2018, signals of strong anomalous meridional winds propagated from the West equatorial Pacific ($100^{o}$E - $120^{o}$E) to the east coast of Africa, with the phase speed ($C_{p}$) of about 12 m/s. These westward propagating signals possess wave characteristics which resemble the equatorially trapped atmospheric waves of synoptic time scales  \cite{b8,b9,b10}. The variability associated with meridional winds corresponding to 12-20 days and 20-60 days at 200hPa, in Fig.~\ref{fig:fig10}(third figure) and Fig.~\ref{fig:fig10}(fourth figure) respectively, do not reveal any wave characteristics in the wind field.

Similarly, Hovemöller diagrams are prepared from meridional winds at 200hPa pressure level in 2019 for the three variability modes (Fig.~\ref{fig:fig11}). As observed in 2018, westward propagating waves of 5-8 days period, resonated the upper troposphere for the second consecutive year (Fig.~\ref{fig:fig11}(first figure)). The high-frequency waves originated near the east Indian Ocean ($100^{o}$E) crossed over Kerala and finally reached Africa, with the same phase speed as in 2018. As anticipated, the other modes had not assumed significance during the extreme event (Fig.~\ref{fig:fig11}(second figure) and Fig.~\ref{fig:fig11}(third figure)). Therefore, the westward-propagating high frequency waves were simultaneously excited in the atmosphere with the anomalous convection over Kerala.

The ERA5 specific humidity in the atmosphere is vertically integrated from near surface (1000 hPa) to the mid-troposphere (600 hPa) and bandpass-filtered for the high-frequency mode for the year 2018, with the objective of identifying the source of moisture (Fig.~\ref{fig:fig12}). The figure is overlaid with the winds (5-8 days bandpassed) at 700 hPa. Alternate cyclonic and anti-cyclonic circulations were observed to travel westward at mid-troposphere from Equatorial Pacific to Arabian Sea, which is nothing but the manifestation of the westward-propagating high-frequency wave train. The waves appear to possess zonal wave length of about 6000 km as they propagate with a phase speed of 12m/s and are trapped to the equator. One observes the transport of moisture along the wave trajectory, especially with the cyclonic vortices, which infers that the mid-tropospheric convection in 2018 was the response to a low pressure region created by the circulations in the field induced by waves. As a consequence, the waves dilated the wind field to stimulate convection and further conveyed additional moisture to cause extreme precipitation on the windward side of Kerala in consecutive years.

\section{Summary \& Conclusions}
Kerala experienced a spate of flood events in 2018 and 2019 during Summer Monsoon. This study resolves to understand the dynamics of anomalous rainfall events in the state and gives better insights from the available observations. The summary of analysis and the conclusions thus drawn are discussed here.

\begin{enumerate}
\item Extreme precipitation events recurred in Kerala in 2018 and 2019, which peaked on $15^{th}$ and $8^{th}$ August respectively. The convective activity at mid-troposphere and upper-troposphere were strongly correlated with the spatial distribution of precipitation.
\item Fourier Transforms performed on the precipitation anomalies exposed the prominent modes of variability -- 5 to 8 days, 12 to 16 days and 20 to 60 days. Out of these, the high frequency mode (5 to 8 days) is observed to dominate over the other two in 2018 during the extreme event, although all the three seemed to possess comparable magnitudes as they fell in-phase in 2019.
\item The high frequency mode of period 5-8 days dominated mid-troposphere in 2018 as well as at the upper-troposphere in 2019, all particular to the period of extreme events in the state. The 12-20 days mode (Quasi-biweekly Oscillation) and 20-60 days mode (Madden-Julian Oscillation) have characterized the zonal winds in 2019, thereby justifying their contribution to the extreme event of the year, but not exclusively. Hence, the 5-8 days high frequency mode is solely responsible for the extreme precipitation in Kerala.
\item The high-frequency mode manifested as westward-propagating high frequency tropical atmospheric waves of characteristic phase speed of nearly 12m/s, which originated near east equatorial Indian Ocean or tropical West Pacific and travelled to the east coast of Africa, and coincide with the same period of extreme rainfall events over Kerala. Moreover, the waves appeared as cyclonic and anti-cyclonic circulations trapped to the equator, which dilated the wind field and transported moisture as it propagated. The waves not only stimulated convection along its trajectory, but also ensured sufficient moisture availability. Therefore, the convective activities which intensified in the mid-troposphere and upper-atmosphere in 2018 and 2019 respectively, were the direct consequence of the equatorially-trapped high frequency waves, which played a significant role in driving the recurrent anomalous precipitation in the South Indian state. 
\end{enumerate}

\section*{Acknowledgment}
The author thanks Indian Meteorological Department (IMD) and Asia-Pacific Data Research Center (APDRC) services for granting free online access to the observations and Reanalysis data for all users.

\section*{Author Contributions}
K.S.R. performed the analysis, drawn conclusions and prepared the manuscript.

\section*{Competing Interests}
The author declares that there exist no competing interest with any person or agency in matters related to this research, and has not received any funds or grants in any form for the same.

\end{document}